\documentclass[a4paper,11pt]{article}
\usepackage{pos}
\usepackage{lineno}
\usepackage{slashed}
\usepackage{subcaption}
\usepackage{mathtools}
\usepackage{commath}
\usepackage{multirow}
\usepackage{hyperref}
\usepackage{cleveref}
\usepackage{booktabs}
\usepackage{nicefrac} 
\usepackage{comment}
\usepackage{graphicx}
\usepackage{float}
\usepackage{enumitem}
\usepackage{physics}

\title{SM fermion scattering off electric charge and CP-violating domain walls in the 2HDM}
\ShortTitle{SM Fermion Scattering off Charge and CP-Violating DW in the 2HDM}

\author*[a]{Mohamed Younes Sassi}
\author[a,b]{Gudrid Moortgat-Pick}

\affiliation[a]{II. Institut für Theoretische Physik,
University of Hamburg,\\Luruper Chaussee 149, 22761 Hamburg, Germany}

\affiliation[b]{Deutsches Elektronen-Synchrotron DESY, Notkestr. 85, 22607 Hamburg, Germany}

\emailAdd{mohamed.younes.sassi@desy.de}
\emailAdd{gudrid.moortgat-pick@desy.de}

\abstract{In several models of beyond Standard Model physics (BSM) discrete symmetries play an important role. For instance, in order to avoid flavor changing neutral currents (FCNC), a discrete $Z_2$ symmetry is imposed on Two-Higgs-Doublet-Models (2HDM). This can lead to the formation of domain walls (DW) as the $Z_2$ symmetry gets spontaneously broken during electroweak symmetry breaking (EWSB) in the early universe.
Due to this simultaneous spontaneous breaking of both the discrete symmetry and the electroweak symmetry, the vacuum manifold has the structure of two disconnected 3-spheres and the formed domain walls can exhibit several special properties in contrast to standard domain walls. We focus on some of these properties such as CP and electric charge violating vacua localized inside the domain walls. The breaking of $U(1)_{em}$ inside the wall leads to the known phenomenon of "clash-of-symmetries" mechanism, meaning that the symmetry group inside the wall is smaller than the symmetry group far from the wall.  We also discuss the scattering of top quarks off such types of domain walls and show, for example, that they can be reflected or transmitted off the wall as a bottom quark.}

\FullConference{The European Physical Society Conference on High Energy Physics (EPS-HEP2023)\\
 21-25 August 2023\\
Hamburg, Germany\\}

\begin{document}

\begin{flushright}
DESY-23-173
\end{flushright}

\maketitle

\section{Introduction}
Two-Higgs-doublet models are a minimal and well motivated extension of the standard model of particle physics. In order to avoid flavor changing neutral currents that occur in such models, one usually imposes a discrete $Z_2$ symmetry on the scalar and fermion sector. The spontaneous breaking of this symmetry in the early universe might lead, however, to the formation of domain walls \cite{Zurek:1985qw, Kibble:1976sj}. It was recently \cite{Law:2021ing, Viatic:2020yme, Battye:2020sxy, Sassi:2023cqp} established that, due to the simultaneous breaking of the discrete symmetry together with the electroweak symmetry, the DW solution in this model can exhibit non-trivial behavior localized inside the wall, such as electrical charge and CP-violating vacua. In this article, we briefly summarize the different possible DW solutions in the 2HDM and discuss the impact of these types of defects on the scattering of top quarks. The results are based on \cite{Sassi:2023cqp}, where these aspects are discussed in details.

\section{Domain wall solutions in the 2HDM}
\begin{figure}[H]
     \centering
     \begin{subfigure}[b]{0.4\textwidth}
         \centering
         \includegraphics[width=\textwidth]{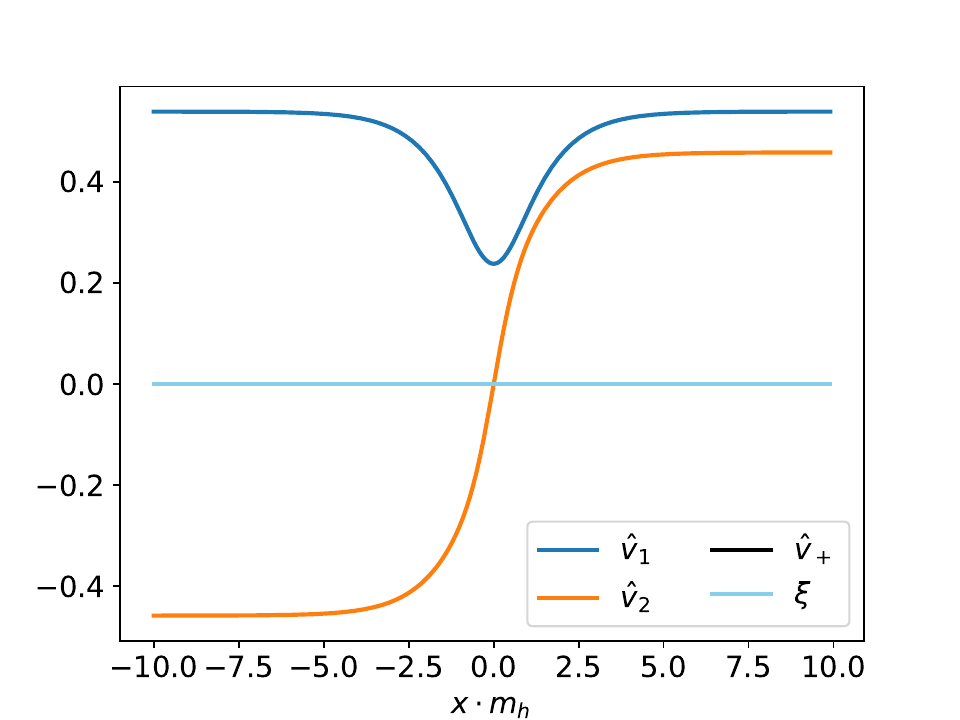}
         \subcaption{Standard DW solution}\label{subfig:dwsolution}
     \end{subfigure}
     \begin{subfigure}[b]{0.4\textwidth}
         \centering
         \includegraphics[width=\textwidth]{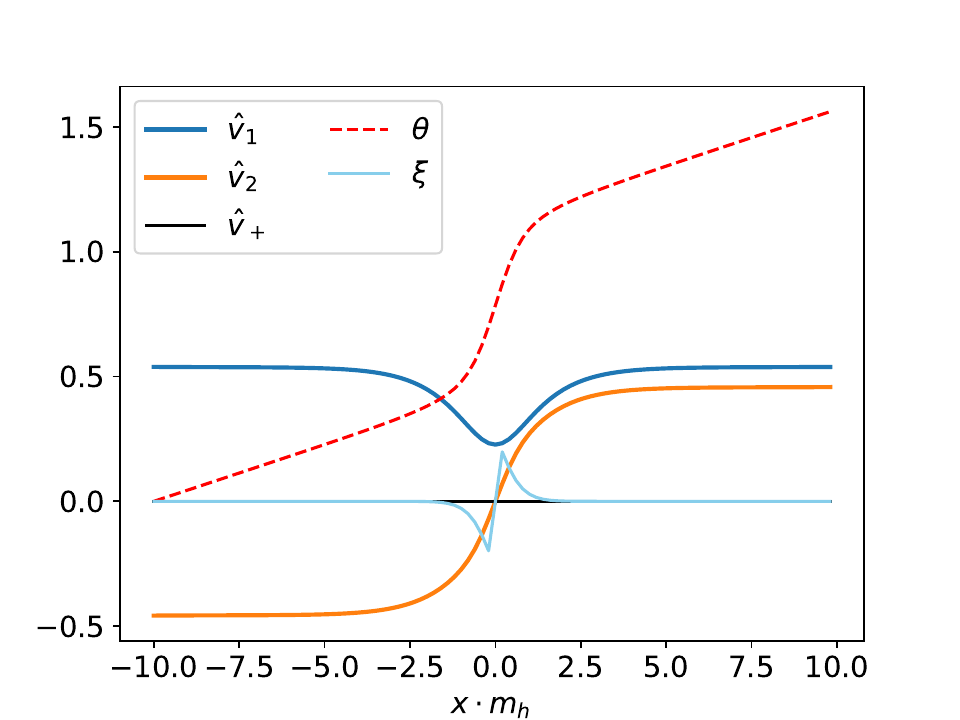}
         \subcaption{Solution with varying $\theta$}\label{subfig:theta}
     \end{subfigure}
    \begin{subfigure}[b]{0.4\textwidth}
         \centering
         \includegraphics[width=\textwidth]{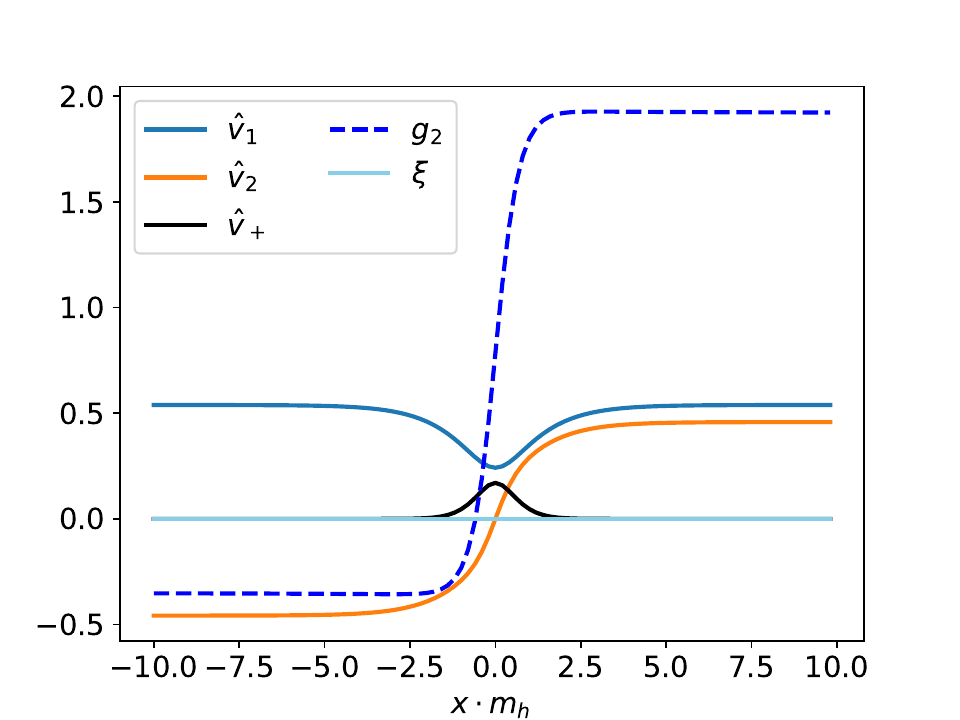}
         \subcaption{Solution with varying $g_2$}\label{subfig:g2}
     \end{subfigure}
\caption{(a) Standard DW solution. (b) DW solution when varying $\theta$ across the domains. (c) DW solution when varying $g_2$. We use the rescaled dimensionless vacuum parameters $\hat{v}_i = v_i/v_{sm} $. The mass parameters are: $m_H = 800 \text{ GeV}$, $m_A = 500 \text{ GeV}$, $m_C = 400 \text{ GeV}$ and $tan(\beta) = 0.85$ \cite{Sassi:2023cqp}. }
\label{fig:stan}
\end{figure}

We begin with briefly introducing the model. In the 2HDM the SM Higgs sector is extended by an extra scalar doublet, i.e. $\Phi_{1,2}$, charged under $SU(2)_L\times U(1)_Y$. We also impose a $Z_2$ symmetry, $\Phi_1 \rightarrow \Phi_1$, $\Phi_2 \rightarrow -\Phi_2 $, in order to avoid FCNCs. The renormalizable potential, invariant under the SM as well as the $Z_2$ symmetries, is given by:
\begin{align}
  \notag V_{\text{2HDM}} &=  m^2_{11}\abs{\Phi_1}^2 + m^2_{22}\abs{\Phi_2}^2 + \frac{\lambda_1}{2}\abs{\Phi_1}^4 + \frac{\lambda_2}{2}\abs{\Phi_2}^4   + \lambda_3\abs{\Phi_1}^2\abs{\Phi_2}^2 \\ & + \lambda_4\bigl(\Phi_1^{\dagger} \Phi_2\bigr)\bigl(\Phi_2^{\dagger} \Phi_1\bigr)  
    +\biggl[\frac{\lambda_5}{2}\bigl(\Phi_1^{\dagger} \Phi_2\bigr)^2 + h.c\biggr].
\end{align}
The 2HDM includes 8 scalar degrees of freedom. In our work we adopt the non-linear representation \cite{Law:2021ing,Battye:2020sxy} to parameterize the vacua: 
\begin{align}
  &   \Phi_1 = U \tilde{\Phi}_1 = U \dfrac{1}{\sqrt{2}}
    \begin{pmatrix}
          0 \\      v_1
     \end{pmatrix},      
&& & \Phi_2 = U \tilde{\Phi}_2 = U \dfrac{1}{\sqrt{2}}
      \begin{pmatrix}
     v_+ \\
     v_2e^{i\xi}
      \end{pmatrix} ,  
\label{eq:nonlinrep2}      
\end{align}
where $U$ is an element of the $\text{SU(2)}_L\times\text{U(1)}_Y$ group that is given by:
\begin{equation}
        \text{U}(x) = e^{i\theta(x)} \text{exp}\biggl(ig_i(x)\sigma_i\biggl),
\label{eq:EWmatrix}        
\end{equation}
where $\theta$, $g_{1,2,3}$ are, respectively, the hypercharge angle and Goldstone modes, $\sigma_i$ denotes the Pauli matrices, i.e. the generators of SU(2). Vacua with $v_+ \neq 0$ are called charge breaking vacua as they lead to the spontaneous breaking of $U_{em}(1)$. For $\xi \neq 0$, the vacuum is CP-violating and for $v_+ = 0$ and $\xi = 0$, the vacuum is neutral. In our work, we only consider neutral vacua at the boundaries $ \pm \infty$.\\
Due to the breaking of the discrete symmetry alongside the electroweak symmetry, the vacuum manifold of the theory is made of two disconnected 3-spheres:   $$ M = (\text{SU(2)}_L \times \text{U(1)}_Y \times Z_2)/\text{ U(1)}_{em} \simeq Z_2 \times S^3. $$
In order to get domain wall solutions we need to have the vacua of the doublets at the boundaries $\pm \infty$ to lie on different disconnected sectors of the vacuum manifold. In contrast to simple models where only the discrete symmetry is spontaneously broken, the breaking of the EW symmetry will lead to a multitude of choices for the vacua at the boundaries. If we suppose that the configuration $\{\Phi_-, \Phi_+\}$ leads to a domain wall solution:
\begin{align}
    \Phi_- = \biggl\{ \dfrac{1}{\sqrt{2}}
    \begin{pmatrix}
          0 \\      v^*_1
     \end{pmatrix}  , \dfrac{1}{\sqrt{2}}
      \begin{pmatrix}
     0 \\
     -v^*_2
      \end{pmatrix}  \biggr\}, && \Phi_+ = \biggl\{ \dfrac{1}{\sqrt{2}}
    \begin{pmatrix}
          0 \\      v^*_1
     \end{pmatrix}  , \dfrac{1}{\sqrt{2}}
      \begin{pmatrix}
     0 \\
     v^*_2
      \end{pmatrix}\biggr\}.
\label{eq:boundaryconditions}      
\end{align}
In that case $\{\Phi_-, U\Phi_+\}$, is also a DW solution. The different choices of the Goldstone modes and the hypercharge angle at the boundaries will lead to the existence of several classes of DW solutions: standard DW, CP-violating DW and electric charge-violating DW.
\begin{figure}[H]
     \centering
     \begin{subfigure}[b]{0.4\textwidth}
         \centering
         \includegraphics[width=\textwidth]{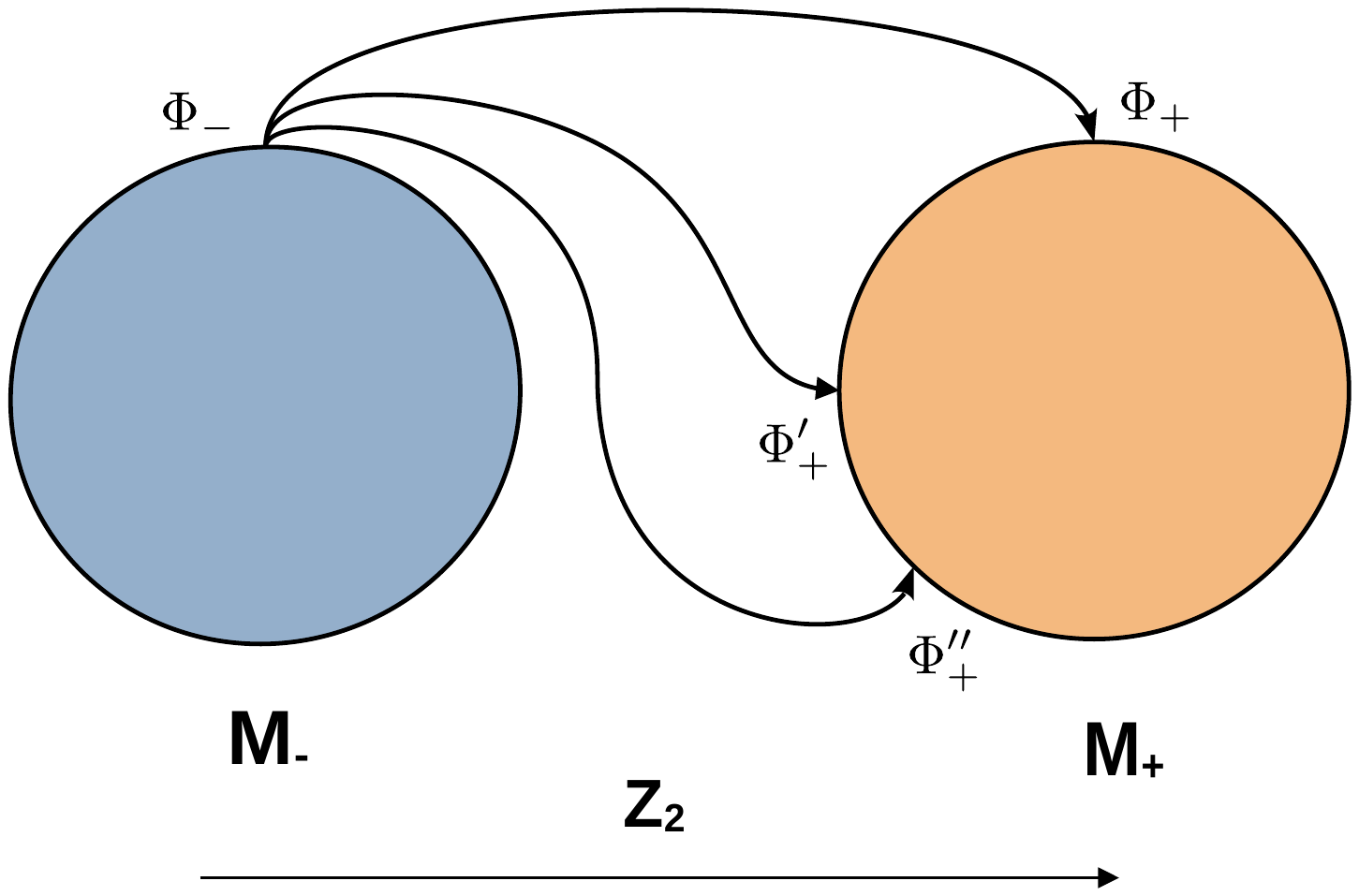}
         \subcaption{Vacuum manifold M of the model.}\label{subfig:vacmanifold}
     \end{subfigure}
     \hfill
     \begin{subfigure}[b]{0.4\textwidth}
         \centering
         \includegraphics[width=\textwidth]{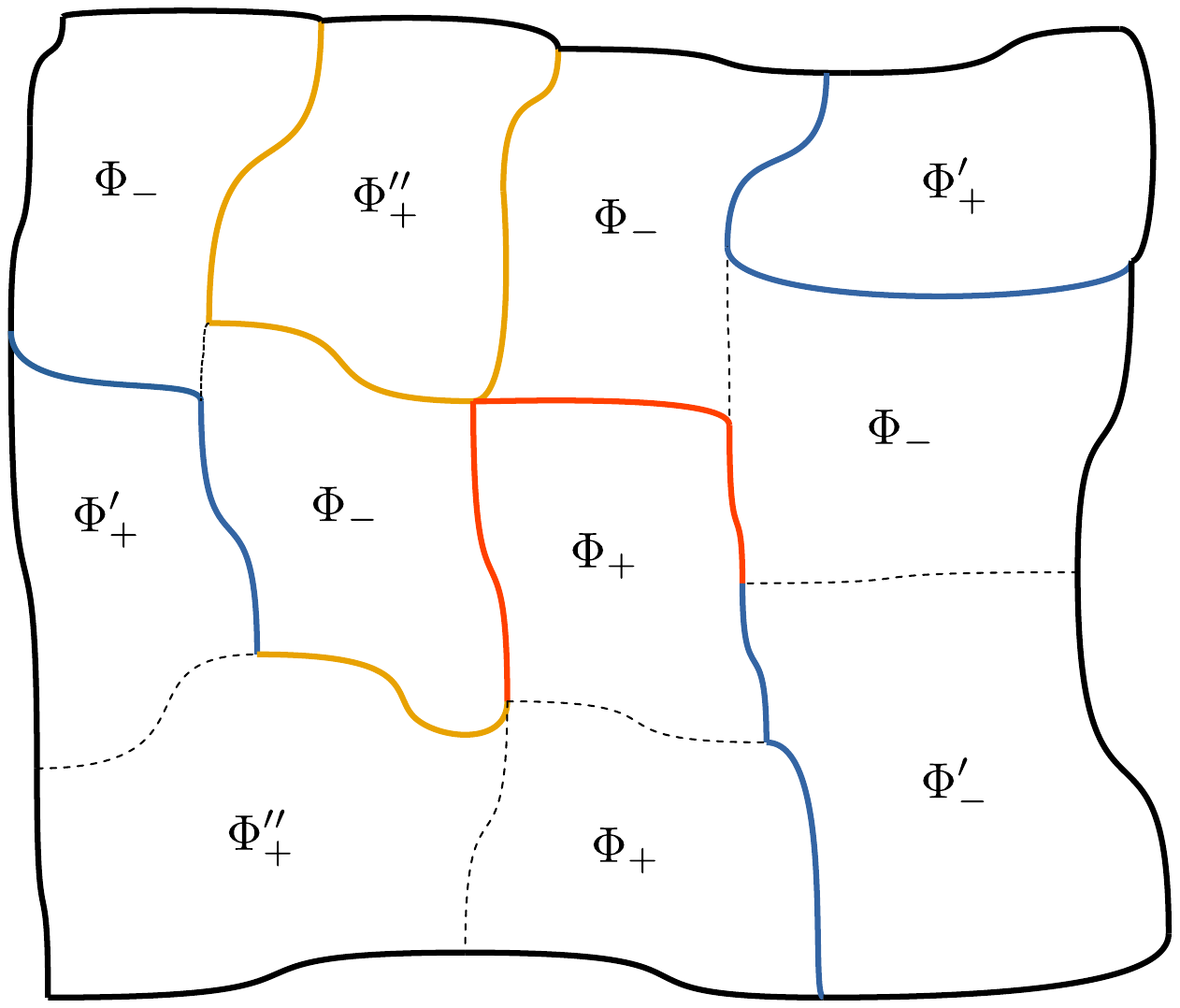}
         \subcaption{A patch of the universe after EWSB.}\label{subfig:earlyuniversevac}
     \end{subfigure}
\caption{(a) Vacuum manifold M of the model. In this case M consists of two disconnected sectors $M_-$ and $M_+$ related by the $Z_2$ symmetry and all the vacua in both sectors are degenerate. The elements of each sector are related by $\text{SU(2)}_L \times \text{U(1)}_Y$ transformations. $\Phi'_+$ and $\Phi''_+$ are related to $\Phi_+$ by different gauge transformations of $\text{SU(2)}_L \times \text{U(1)}_Y$.
(b) After EWSB, causally disconnected regions of the universe can end up in different vacua of the vacuum manifold. Regions that end up with vacua in separate sectors of M can have different classes of domain walls depending on the Goldstone angles they acquire.}
\label{fig:vacuummanifold}
\end{figure}
\noindent
In order to get a one-dimensional kink solution to a given vacuum configuration where the vacua lie on different sectors of the vacuum manifold, we need to find the lowest energy solution of such a configuration. The energy functional related to the scalar sector of the theory is given by:
\begin{align}
  \notag  \mathcal{E}(x) &= \dfrac{d\tilde{\Phi}^{\dag}_1}{dx} \dfrac{d\tilde{\Phi}_1}{dx} + \dfrac{d\tilde{\Phi}^{\dag}_2}{dx} \dfrac{d\tilde{\Phi}_2}{dx} + \biggl(\dfrac{d\tilde{\Phi}^{\dag}_{1,2}}{dx} U^{\dag}(x)\dfrac{dU}{dx}\tilde{\Phi}_{1,2}(x) + \text{h.c}\biggl) + \tilde{\Phi}^{\dag}_{1,2}(x)\dfrac{dU^{\dag}}{dx}\dfrac{dU}{dx}\tilde{\Phi}_{1,2}(x) \\
    & + V_{2HDM}(\Phi_1, \Phi_2).
\label{eq:energydensity}    
\end{align}
This minimization leads to equations of motion for the 8 vacuum parameters. In this letter we only discuss the special cases that occur when having only a single Goldstone mode being different at the two boundaries. For a comprehensive discussion of the complete theory, we refer the reader to \cite{Sassi:2023cqp} for more details. We therefore can simplify the matrix $U$ in (\ref{eq:EWmatrix}) and end up with 5 degrees of freedom. \\
We start by considering the case when the two vacua of the domains at $\pm \infty$ have the same values for $\theta$ and $g_i$ leading to $U(x) = const$. In this case, the energy functional (\ref{eq:energydensity}) reduces to:
\begin{align}
  \notag  \mathcal{E}_s(x) &= \frac{1}{2} (\frac{dv_1}{dx})^2 +  \frac{1}{2} (\frac{dv_2}{dx})^2 + \frac{1}{2} (\frac{dv_+}{dx})^2 + \frac{1}{2}v^2_2(x)(\frac{d\xi}{dx})^2 + V_{2HDM}. 
\label{eq:ensta}  
\end{align}
The equations of motion of the vacuum parameters $v_i(x)$ and $\xi(x)$ with the boundary conditions (\ref{eq:boundaryconditions}) are solved numerically using the gradient flow method \cite{Viatic:2020yme,Battye:2011jj} and the results are shown in Figure \ref{subfig:dwsolution}. Notice that the profile of the vacuum $v_1(x)$ also changes inside the DW as both $\Phi_1$ and $\Phi_2$ are coupled via the terms in the potential.
\noindent
We now turn to the case when only the hypercharge angle $\theta$ is different on both domains. In this case we obtain several extra contributions to (\ref{eq:energydensity}):
\begin{align}
  \notag  \mathcal{E}(x) &= \mathcal{E}_s(x) +  \dfrac{1}{2}v^2_1(x)\biggl(\dfrac{d\theta}{dx}\biggr)^2  + \dfrac{1}{2}v^2_2(x)\biggl[  \biggl(\dfrac{d\theta}{dx}\biggr)^2 + 2\dfrac{d\theta}{dx}\dfrac{d\xi}{dx} \biggr]  + \dfrac{1}{2}v^2_+(x)  \biggl(\dfrac{d\theta}{dx}\biggr)^2. 
\end{align}
In Figure \ref{subfig:theta}, we show the numerical results of the vacuum configuration for initial values of the hypercharge $\theta(-\infty) = 0$ and $\theta(+\infty) = \pi/2$ using Dirichlet boundary conditions. We observe that, as the value for $\theta(x)$ changes across both domains, the value of $\xi(x)$ representing the phase between the two Higgs doublets is non-zero inside the DW. This CP-violating DW solution is, however, unstable as it has a higher energy than the standard DW solution and eventually decays to a solution with $\xi(x) = 0$ and $\theta(x) = const$ for all $x$.\\
We now discuss the DW solution for two domains with different values of $g_2$. In this case, the matrix $U(x)$ simplifies to:
\begin{equation}
    U(x) = \cos\bigl(g_2(x)/2\bigr)I_2 + i\sin\bigl(g_2(x)/2\bigr)\sigma_2 = \begin{pmatrix}
\cos\bigl(g_2(x)/2\bigr) & \sin\bigl(g_2(x)/2\bigr) \\
-\sin\bigl(g_2(x)/2\bigr) & 
 \cos\bigl(g_2(x)/2\bigr) 
\end{pmatrix}.
\end{equation}
The energy functional $\mathcal{E}(x)$ (\ref{eq:energydensity}) is given by:
\begin{align}
   \notag  \mathcal{E}(x) &= \mathcal{E}_s(x) + \frac{1}{8}(v^2_+ + v^2_1 + v^2_2)\biggl(\frac{dg_2}{dx}\biggr)^2 + \frac{1}{2}\frac{dg_2}{dx}\biggl(-v_+\text{cos}(\xi)\frac{dv_2}{dx} + v_2\text{cos}(\xi)\frac{dv_+}{dx} + v_2v_+\text{sin}(\xi)\frac{d\xi}{dx} \biggr).  
\label{eq:potentialg2}   
\end{align}
The numerical results of the equations of motion for the vacuum configuration are given in Figure \ref{subfig:g2}, where we used von Neumann boundary condition due to the fact that the difference in $g_2$ between both domains grows for the solution with least energy. We find that $v_+(x)$ acquires a condensate inside the DW. This leads to the breaking of the electromagnetic symmetry $U(1)_{em}$. Such a behavior can be understood in the framework of the "clash-of-symmetries" phenomenon introduced in \cite{Davidson:2002eu, Davidson:2007cf, Shin:2003xy}, which leads to the fact that the symmetry inside the DW is smaller than the symmetry outside the DW in case the non-broken subgroups $U^+(1)_{em}$ and $U^-(1)_{em}$  on the asymptotic boundaries outside the wall are embedded differently inside $SU(2)_L\times U(1)_Y$. 

\section{Scattering of top quarks off the domain walls}
\begin{figure}[H]
     \centering
     \begin{subfigure}[b]{0.35\textwidth}
         \centering
         \includegraphics[width=\textwidth]{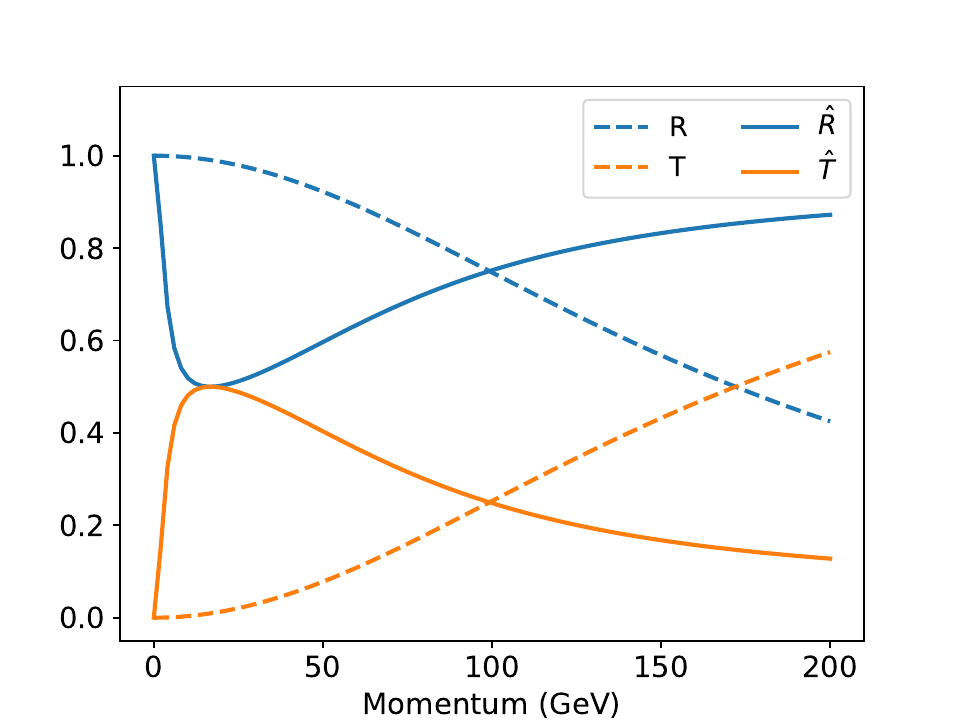}
         \subcaption{CP-violating scattering}\label{subfig:cpscattering}
     \end{subfigure}
     \begin{subfigure}[b]{0.35\textwidth}
         \centering
         \includegraphics[width=\textwidth]{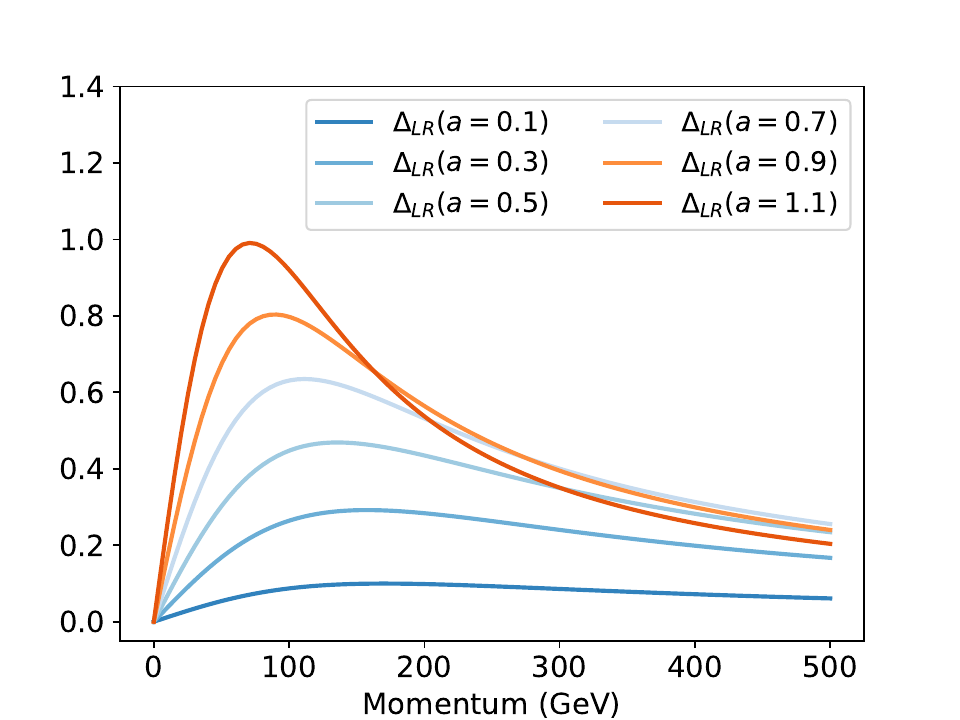}
         \subcaption{$\Delta_{LR}=R_L-R_R$ for different $\tilde{v}_2$}\label{subfig:deltarl}
     \end{subfigure}
    \begin{subfigure}[b]{0.35\textwidth}
         \centering
         \includegraphics[width=\textwidth]{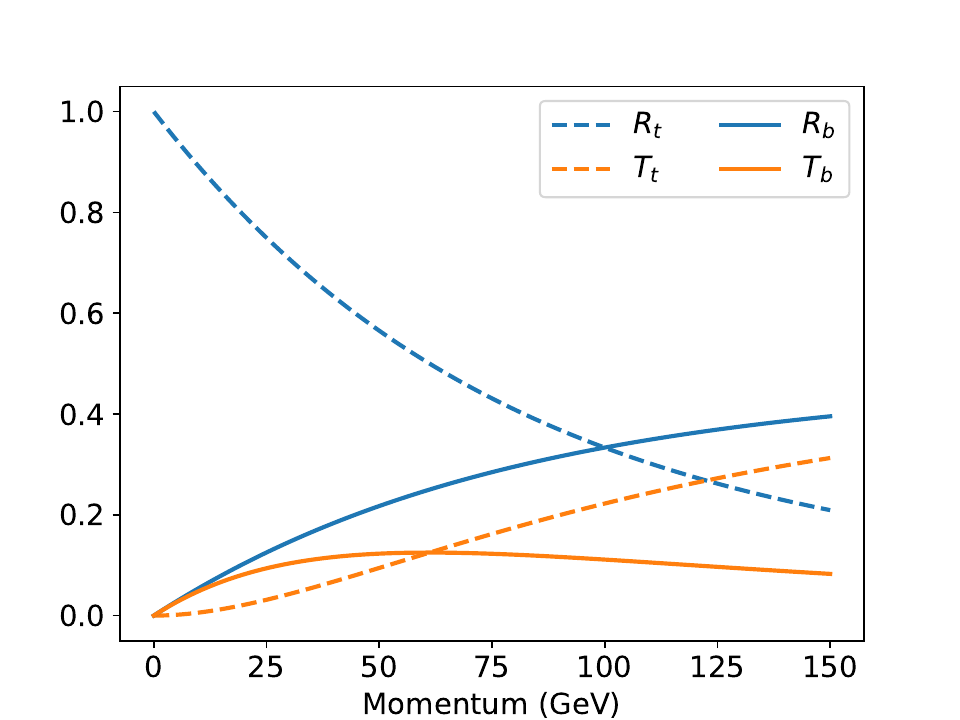}
         \subcaption{Electric charge violating scattering}\label{subfig:Charge}
     \end{subfigure}
    \begin{subfigure}[b]{0.35\textwidth}
         \centering
         \includegraphics[width=\textwidth]{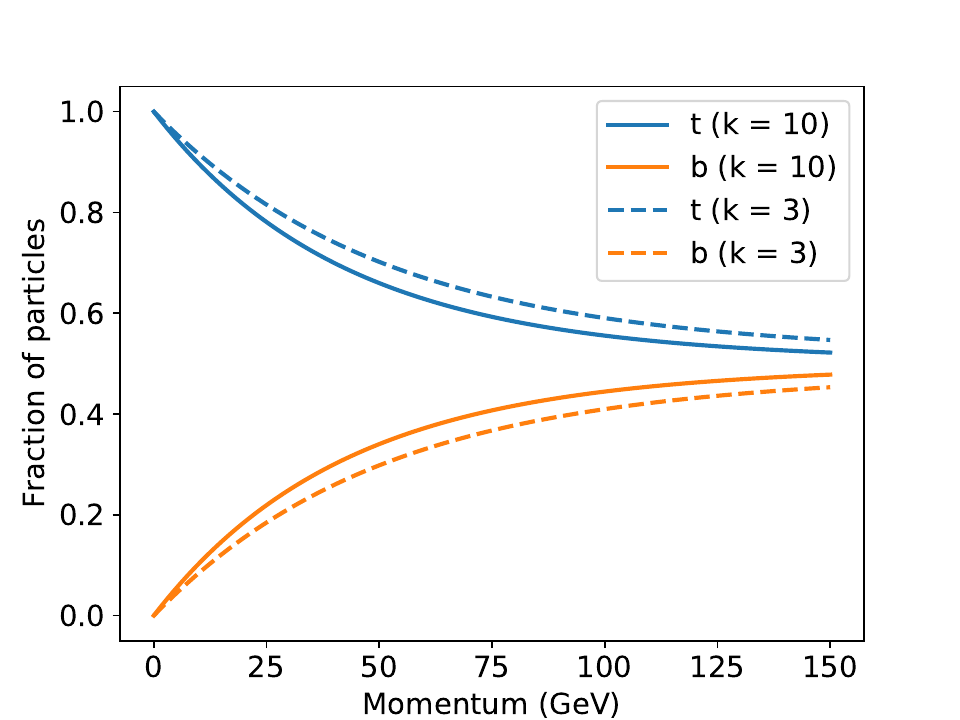}
         \subcaption{Type of particle after scattering}\label{subfig:fractioncharge}
     \end{subfigure}
\caption{(a) Reflection and transmission coefficients for the scattering off a CP-violating DW (solid) and standard DW (dashed) for $a = 2\hat{\tilde{v}}_2 = 2.1$ and $\Delta \theta = 1$. (b) Difference in the reflection coefficient for left ($R_L$) and right-handed ($R_R$) particles for several choices of the CP-violating measure $a = 2\hat{\tilde{v}}_2$. (c) Reflection and transmission rates as a bottom and top quark after the scattering of an incident top quark with the electric charge violating DW, we use $v_+ = 10 \text{ GeV}$ and $\Delta g_2 = 1$. (d) Fraction of particles after the electric charge breaking scattering as a function of the momentum of the incident top quark for different values of $k=y_t\hat{v}_+$.}
\label{fig:fermionscattering}
\end{figure}

In this section we discuss the scattering of top quarks off the different types of domain walls in the 2HDM. To do this, we solve the Dirac equation of the top quark in the background of a Higgs field with the profile of a domain wall.\\
We consider the type-two 2HDM, where the top quark couples to the second doublet while the bottom quark couples to the first doublet. In order to get analytical results, we use the thin-wall approximation which simplifies the spatial profile of the vacuum considerably:
\begin{align}
    v_2(x) = -\hat{v}_2\Theta(x) + \hat{v}_2\Theta(-x).
\end{align}
As for the vacua $v_1(x)$, $v_+(x)$ and $\tilde{v}_2(x)=\text{Im}(v_2(x)e^{i\xi(x)})$, it is possible to approximate them with a delta distribution:
\begin{align}    
    v_1(x) = v_1 + \hat{v}_1 \delta(x), &&
    v_+(x) = \hat{v}_+ \delta(x), &&
    \tilde{v}_2(x) =  \hat{\tilde{v}}_2 \delta(x-\epsilon) +  \hat{\tilde{v}}_2 \delta(x+\epsilon), \label{eq:v2tilde}   
\end{align}
where $\hat{v}_{1,2,+}$ are dimensionless parameters defined as $\hat{v}_{1,2,+} = v_{1,2,+}/ \text{GeV}$.
For the CP-violating case, the Dirac equation of the top quark is given by:
\begin{align}
    \biggl( i\slashed{\partial} - \slashed{\partial}\theta(x)(Y_{t,L}P_L + Y_{t,R}P_R) - m_{R}(x) -im_I(x)\gamma_5   \biggr)t(x,t) = 0,
\label{eq:diraccp}    
\end{align}
where $m_R(x) = y_uv_2(x)\cos(\xi(x)) \approx y_uv_2(x)$ and $m_I(x) = y_uv_2(x)\sin(\xi(x)) \approx y_uv_2(x)\xi(x) = y_u\tilde{v}_2(x)$ and $t(x,t)$ representing the top quark spinor.
The Dirac equation is solved using a plane-wave ansatz. After matching the results for $x>0$ and $x<0$ and getting the solution of the top quark spinor, we compute the reflection and transmission coefficients of the top quark off the wall using $\hat{\text{R}}\text{(p)} = -\frac{\mathcal{J}_{ref}}{\mathcal{J}_{inc}} $ and $\hat{\text{T}}\text{(p)} = \frac{\mathcal{J}_{tra}}{\mathcal{J}_{inc}}$, where $\mathcal{J}$ denotes the currents of the reflected, transmitted and incident particles. The results are shown in Figure \ref{subfig:cpscattering} where we also show the reflection and transmission coefficients for the scattering of top quarks off a standard DW (dashed lines). We also verify that left-handed and right-handed particles are reflected off the wall with different rates (see Figure \ref{subfig:deltarl}).
\noindent
We now discuss the case when $v_+(x) \neq 0$ inside the DW. The Dirac equations for the top and bottom quarks are coupled due to the non-zero $v_+$:
\begin{align}
   & i\slashed{\partial}b(x,t) + \frac{i}{2}\biggl(\slashed{\partial}_xg_2(x)\biggr)P_Lt(x,t) - y_d v_1(x)b(x,t) + y_uv_+(x)P_Rt(x,t)  = 0, 
\label{eq:diraccharged1}   
   \\ &
   i\slashed{\partial}t(x,t) - \frac{i}{2}\biggl(\slashed{\partial}_xg_2(x)\biggr)P_Lb(x,t) - y_u v_2(x)t(x,t) + y_uv_+(x)P_Lb(x,t)  = 0, 
\label{eq:diraccharged2}   
\end{align}
which leads to the possibility that the top quark scatters as a bottom quark after its interaction with the DW. The results for the reflection and transmission coefficients as top or bottom quarks are shown in Figure \ref{subfig:Charge}. We find that increasing the momentum of the incident top quark leads to a higher probability for conversion into a bottom quark (see Figure \ref{subfig:fractioncharge}). Another feature of this scattering is that all the produced bottom quarks are left-handed.

\section{Conclusions}
In this proceeding we summarized our discussion of domain walls in the 2HDM and their interaction with top quarks \cite{Sassi:2023cqp}. We showed that, due to the fact that the vacuum manifold has a non-trivial disconnected structure, we obtain different classes of domain wall solutions with different properties inside the core of the wall such as CP or electric charge violating vacua. We also showed that the scattering of top quarks off these types of domain wall occurs in a CP and electric charge violating manner. Consequences of these phenomena on early universe cosmology are subject of future work.

\subsection*{Acknowledgments}
This work is funded by the Deutsche Forschungsgemeinschaft (DFG) through Germany’s Excellence Strategy – EXC 2121 “Quantum Universe” — 390833306.

\bibliographystyle{JHEP}
\bibliography{refrences}

\end{document}